# Ultra-low Power Nanoelectromechanical Memory Based on Location-controllable Nanogap System

Jian Zhang,[†,‡] Ya Deng,[†,‡] Xiao Hu,[†,‡] Jean Pierre Nshimiyimana,[†,‡] Siyu Liu,[†] Xiannian Chi,[†,‡] Pei Wu,[†,‡] Fengliang Dong,[†] Peipei Chen,[†] Weiguo Chu,[*,†] Haiqing Zhou,[*,§] and Lianfeng Sun[*,†]

[†]CAS Key Laboratory of Nanosystem and Hierarchical Fabrication, Nanofabrication laboratory, CAS Center for Excellence in Nanoscience, National Center for Nanoscience and Technology, Beijing 100190, China.

[§]Department of Physics and TcSUH, University of Houston, Houston, Texas 77204, United States

[‡]University of Chinese Academy of Sciences, Beijing 100049, China.

ABSTRACT: Nanogap engineering of low-dimensional nanomaterials, has received considerable interest in a variety of fields, ranging from molecular electronics to memories. Creating nanogaps at a certain position is of vital importance for the repeatable fabrication of the devices. In this work, we report a rational design of non-volatile memories based on sub-5 nm nanogaped single-walled carbon nanotubes (SWNTs) via the electromechanical motion. The nanogaps are readily realized by electroburning in a partially suspended SWNTs device with nanoscale region. The SWNT memory devices are applicable for both metallic and semiconducting SWNTs, resolving the challenge of separation of semiconducting SWNTs from metallic ones. Meanwhile, the memory devices exhibit excellent performance: ultra-low writing energy ($4.1 \times 10^{-19}$ J per bit), ON/OFF ratio of $10^5$, stable switching ON operations and over 30 hours retention time in ambient conditions, which are of great potential for applications.

Continuation of Moore's law to the sub-10-nm scale requires the development of new technologies for creating electrode nanogaps, in architectures which allow a third electrostatic gate.[1-3] Nanogap engineering of low-dimensional nanomaterials, have the potential to fulfill this need, provided their structures and properties at the moment of gap formation could be controlled, which has been of emerging interest in a variety of fields, ranging from molecular electronics to memories.[2, 4-9] Nanogaps also have wide applications in nanoelectromechanical switching (NEMS), where electrostatic forces are used to mechanically deflect an active element into physical contact with an electrode, thus changing the state of the device.[10-11] Single-walled carbon nanotubes (SWNTs) are one-dimensional hollow structure, rolled by one-atom-thick graphene, which is a more promising material to act as the active element in NEMS.[12-15] Meanwhile, the ultimate limit for the miniaturization of electronics is the nanoscale in three dimensions, so restricting the deflection of SWNTs in the nanogaps can achieve this goal.

Recently, graphene nanogaps are created by feedback-controlled electroburning at the position of hundreds of nanometers-wide constrictions, which seems to be more reliable and precise than other approaches.[2, 4, 16] Even so, it is not applicable for creating nanogaps in carbon nanotubes using electroburning method, since it is difficult to control their positions due to the random nature of the electroburning process. Alternatively, some other strategies are introduced to creat nanogaps in carbon nanotubes, such as electron beam lithography and atomic force microscopy,[5, 7, 17] which can precisely control the position of the nanogaps. However, there are some disadvantages restricting their applications, such as the contamination by organic resist residues and the low yield of nanogaps. Thus, controlling the creation of the carbon nanotube nanogap at a certain position is a critical step to fabricate the functional molecular devices.

In this work, we introduce a novel design for fabricating a switching memory in a SWNT device by the electroburning method. The nanoscale gap is created in the suspended section of SWNT nearby the electrode using the electroburning method. Then the end of SWNT can be readily deflected in this nanogap by electrostatic forces to achieve the switching operations. The basic switching operations (OFF to ON, nonvolatility and ON to OFF), the endurance test and retention time test of the memory devices are investigated in ambient conditions. The power consumption and operation speed are calculated and found superior to commercially available flash memories,[18-20] suggesting great potential applications.

RESULTS AND DISSCUSSION

The SWNTs used in this work were grown by floating catalytic chemical vapor deposition (See Methods).[21] Then electron-beam lithography was utilized to fabricate the suspended SWNT devices (See Methods and Supporting Information S1). A typical false-color SEM image of a SWNT device with a tilted angle of 60° is shown in Figure 1a. An individual SWNT is connected to two nickel electrodes: source (S) and drain (D). Two side gates (G1 and G2) are patterned at the nearby location of source and drain electrodes, respectively. The distance between the gate electrodes and the SWNT is about 200 nm. There is a small angle ($\alpha \sim 10°$) between the SWNT and source (drain) electrode.

In the lift-off process, the devices were soaked in acetone and ethanol successively to remove the PMMA. The evaporation of ethanol on the devices generates a capillary force that pulls down the SWNT to adhere on the silicon substrate partially.[22] Figure 1b shows the typical atomic force microscopy (AFM) topographic image of a SWNT device with tapping mode. The height profile along the SWNT is shown in Figure 1c. These results indicate that there are five sections in the as-prepared SWNT devices: embedded section in the source electrode (section i); suspended and oblique section (Section ii); adhered to the substrate (section iii); suspended and oblique section (section iv); embedded section in the drain electrode (section v). The height ($h$) of the suspended section (ii and iv) is 60~100 nm. Sections i, ii, and iii of the SWNT device are shown in the schematic diagram in Figure 1d. The average value of the projected lengths of the suspended SWNT (section ii and iv) is about 226 nm (Supporting Information S2). When there is an angle of 10 ° between the SWNTs and the electrodes, the maximum distance between the suspended SWNTs and the electrodes is below 40 nm, as shown in Figure S2b. A sub-40 nm-wide region is created between the suspended section of the SWNT and the electrodes, which provides a reliable confinement to create a nanogap by electroburning method.

The low resistance ($R \approx 140$ k$\Omega$) and linear relationship between drain current ($I_D$) and drain voltage ($V_{DS}$) indicate Ohmic contact between the electrodes and the SWNT as shown in Figure 1e. To have a better understanding about the electrical properties of this SWNT device, the drain current ($I_D$) versus gate voltage ($V_G$) are measured at different gate voltages in ambient conditions.[1, 23-24] In all devices we have measured (~89), there are two types of SWNTs: semiconducting and metallic SWNTs as indicated by the $V_G$ modulation of the $I_D$. Figure 1f shows the typical modulation curve of a metallic SWNT device. The $V_G$ was applied at the backgate of doped silicon, and the thickness of the SiO$_2$ layer is 300 nm. The typical modulation curve of semiconducting SWNT device is shown in Supporting Information Figure S3a. In previous studies using cantilevered carbon nanotube devices, multiwalled carbon nanotubes rather than SWNTs are often used because they are metallic and the SWNTs are usually a mixture of metallic and semiconducting nanotubes.[12] As shown later in this work, the SWNT memory devices reported in this work are applicable both for metallic or semiconducting SWNTs, resolving the challenge of separation of semiconducting SWNTs from metallic ones.[25-26]

The switching memory devices were fabricated by electroburning method that relies on Joule heating to create nanogaps in the suspended section of SWNTs. This method has previously been used to create nanoscale gaps in multiwalled carbon nanotubes and graphene.[2, 4, 10-11, 16] Due to the random nature of the electroburning process, if all part of the SWNT is suspended or adhered on the silicon substrate, the position of the gap in the SWNT is not well controlled. However, if the SWNT in the device is suspended partially as displayed in Figure 1a-d, the situation is quite different. Since the thermal conductivity of air is quite smaller than silicon dioxide, the Joule heating is more liable to congregate at the suspended section of the SWNT.[27-29] Thus, the nanoscale gaps can be controllably created adjacent to the electrodes at a sub-40 nm wide region (section ii and iv in Figure 1b-d). Although it is difficult to control the region where electrical breakdown happens, either in section ii or section iv, the designs of two gate electrodes G1 and G2 make up for this shortcoming. Figure 2a shows the typical *I-V* curve during electroburning of a metallic SWNT device in ambient conditions. When the $V_{DS}$ is swept up with a step of 100 μV, the $I_D$ is monitored. At a bias of about 4.6 V, a very sharp decrease of current from 24 μA to several pA is observed, indicating the occurrence of electrical breakdown of SWNT in the suspended section. As soon as the current dropped, we interrupted the $V_{DS}$ rapidly to prevent the gap from being larger. Figure 2b shows the AFM characterizations of the SWNT device before and after electroburning process. After electroburning, the height of the suspended section (ii) of the SWNT device becomes smaller as shown in the height profile along the SWNT (Figure 2c). This indicates that the SWNT has been broken down, thus forming a nanogap between the end of the SWNT and the source electrode as shown schematically in the inset of Figure 2c. The typical *I-V* curve during electroburning of a semiconducting SWNT device is shown in Supporting Information Figure S3b. A total of 89 SWNT devices (including 58 metallic SWNTs and 31 semiconducting SWNTs) were processed by the electroburning method, and the histograms of the electrical breakdown voltage of these devices are shown in Figure 2d. The metallic SWNT devices (grey) have smaller breakdown voltages mainly in the range of 2~10 V, while the semiconducting SWNT devices (red) have larger breakdown voltages mainly ranging from 6 V to 20 V.

We estimate the size of the nanogaps by fitting the measured *I-V* traces to the Simmons model.[16, 30] Typically, the fitting parameters include gap size (*d*) and barrier height ($\Phi_b$). The gap size is derived from the best fitting parameter. As shown in Figure 2e, the red line is the fitted curve with the Simmons model to the *I-V* trace of the SWNT nanogap. The values deduced from this fitting are as followings: gap size *d*=1.94 nm and gap barrier $\Phi_b$=0.5 eV. The details of the Simmons fitting procedure are shown in supporting Information S4. In our devices, the gap size are ranging from 1 nm to 5 nm.

After a nanogap is formed using electroburning method, we can set the device back to its low resistance state by sweeping the voltage past a threshold voltage in ambient conditions. When a voltage is applied to the source electrodes, there is an electrostatic force between the SWNT and the source electrode. This electrostatic force is balanced by the drag forces (including elastic restoring force of SWNT and the adhesive force between the SWNT and the substrate). When a critical voltage is reached, the electrostatic force overwhelms the drag force, which drives the end of the SWNT to move towards the source electrode. Eventually, the end of the SWNT comes into physical contact with the source electrode, which leads to a sharp rise in the current through the device. These processes can be seen clearly form Figure 3a. At the beginning of the sweeping of voltage, the resistance of the SWNT is high (~ GΩ), meaning that the device is in the state of OFF. When the voltage is swept up to about 1.5 V, the current begins to increase quickly. At the voltage about 1.7 V, the current increases sharply and reaches the compliance value of 1.0 μA. When the voltage is swept back to zero, the resistance of the device is low (~600 KΩ) and an Ohmic behavior is seen (Figure 3a), indicating a state of ON. These results demonstrate that the device can act as a memory device and the states can be switched into ON from OFF by a switching voltage of 1.7 V, which is lower than the bias used in conventional dynamic random access memory.[12] The ON (solid) and OFF (translucent) states of the switching device are schematically shown in Figure 3b. The typical *I-V* characteristics of a semiconducting SWNT memory device switching from OFF to ON state is shown in Supporting Information Figure S3c.

It is interesting to note that the ON state of the SWNT memory device is nonvolatile. As shown in Figure 3c, after the SWNT device is switched into ON state (marked with "1"), this state can be kept as indicated by the successive sweeping marked with "2" and "3". This means that the restoring force in the deflected SWNT is smaller than the adhesive force at the contact between the SWNT and the electrode, holding the device at the ON state even when the electrical bias is fully removed.

Another interesting and important question is how to switch the ON state of the memory into OFF state. Actually, applying a voltage to the gate electrode (G1) can switch the memory device to OFF state as shown in Figure 3d. Here a fixed bias of 5 mV is applied between the drain and the source electrodes and the corresponding current ($I_{DS}$) is monitored as the voltage on the gate electrode (G1) is swept up. It can be seen that when the gate voltage is applied, the current ($I_D$) stabilizes at around 5 nA and drops sharply to several pA at the gate voltage of around 21V (switching OFF voltage), indicating the memory device switched into OFF state. This result can be well explained as follows: when the gate voltage is increased, the resultant attractive electrostatic force between the gate electrode and the SWNT increases. At a critical value of voltage, the attractive forces are sufficient to overcome the adhesive force between the SWNT and the source electrode. Then the SWNT is deflected away from the source electrode and a physical separation occurs between them. This makes the SWNT memory device switch into OFF state. The typical *I-V* characteristics of a semiconducting SWNT memory device switching from ON to OFF state is shown in Supporting Information Figure S3d.

It is important to point out that the memory devices reported in this work can switch between the ON and OFF states multiple times in ambient conditions. As shown in Figure 4a, the device can be switched from OFF to ON state by sweeping the source voltage beyond the threshold voltage and switched back from ON to OFF state by sweeping the gate (G1) voltage beyond the threshold voltage. We can repeat SET (from OFF to ON) and RESET (from ON to OFF) multiple times. The values of drain current ($I_D$) are recorded at the drain voltage ($V_{SD}$) of 500 mV. The ON/OFF ratio of the memory device is about $10^5$. It is noted that our memory devices have stable SET operations as shown in the Supporting Information S5. For these 10 times SET operations, the switching ON voltages keep at 1.5-1.8 V, showing stable SET operations of the memory devices. Due to the diversities of different devices, such as the size of the nanogaps and the surface morphologies of the electrodes, the switching ON voltage of our devices are different from each other. Figure 4b is a histogram of switching ON voltage for 53 devices. However, due to the large distance (~200 nm) between the SWNTs and the side gate electrodes, the RESET operations need larger switching voltage compared to SET operations and 12 devices among these 53 devices executed RESET operations successfully. Figure 4c shows the histogram of the switching OFF voltage for these 12 devices.

The retention time of SWNT memory was studied in ambient conditions. Figure 5a shows the *I-V* curves of the SWNT memory at different duration time. Figure 5b is the relevant evolution of the resistance depending on the duration time, corresponding to the *I-V* curves in (a). As the time increases from the beginning to 30 hours, the *I-V* curves show a tendency of increasing resistance. When it is over 40 hours, the SWNT device switches into a high resistance state and becomes OFF state. Its state can be switched into ON by applying a voltage to the source electrode. As shown in Figure 5c, the current is very small (~ pA) at first, and when the voltage is swept to about 4.0 V, the current increases sharply to the value of compliance of 1.0 μA and the device turns into ON state (indicated by sweep "1"). The SWNT device shows nonvolatility as indicated by the successive *I-V* curves marked with "2" and "3". The switching ON voltage of 4.0 V is larger than its pristine value of ~1.5 V, indicating the increase of the size of the gap. And the ON state indicated by sweep "1" has a higher resistance but in the following sweeps (marked with "2" and "3"), its resistance restores to its original value gradually, which may be due to the adsorption at the ends of SWNT.

CONCLUSION

In summary, we have demonstrated a switching memory device consisting of a sub-5 nm gap between the SWNT and the electrode. The switching operations of the memory devices can be achieved by the deflection of the SWNT in this nanogap. The memory devices have the following characteristics: (1) The sub-5 nm gaps between the SWNTs and the electrodes are created by electroburning in a partially suspended SWNTs device with nanoscale region. (2) The SWNT memory devices are applicable for both metallic and semiconducting SWNTs, resolving the challenge of separation of semiconducting SWNTs from metallic ones. (3) The memory devices have ON/OFF ratio of $10^5$, stable SET operations and over 30 hours retention time in ambient conditions. (4) The writing energy of the SWNT memory is as low as $4.1 \times 10^{-19}$ J per bit (the detail is shown in Methods), which is 3~5 orders of magnitude lower than the traditional memory technologies.[18-19] (5) The operation speed of the memory device is dependent on the speed of the cantilever switch and the switching frequency is calculated to be 880 MHz (See Methods).

METHODS

**SWNT Growth.** The SWNTs used in this work were grown by floating catalytic chemical vapor deposition.[21] The experimental setup consisted of a quartz tube with two temperature zones: the first was used for sublimation of the calalysts (60 ℃) and the second was used for SWNT growth (1100 ℃). The catalyst (ferrocene/sulfur powder, molar ration 16:1) was sublimed in the first temperature zone and subsequently carried into the SWNT growth zone by a mixture of 1000 sccm argon and 10 sccm methane. Under the optimized conditions, individual and isolated SWNTs were deposited onto a silicon substrate placed at the collection location, which was at the end part of the quartz tube with temperature about 150 ℃.

**Device Fabrication and Measurement.** In order to fabricate a device with a suspended SWNT, a layer of polymethylmethacrylate (PMMA, thickness of 100 nm) was used as a sacrificial layer and spin-coated on a highly-doped silicon substrate with a 300 nm-thick $SiO_2$ layer. This substrate was placed in the SWNT collection position in the deposition zone of the quartz tube for 5 seconds. Straight and isolated SWNTs were selected and their positions were recorded with Nova Nano SEM 430. Then another PMMA layer (230 nm) was spin-coated on the surface of the sample resulting in the SWNT being sandwiched between the two PMMA layers. Four electrodes (S, D, G1 and G2) were patterned according to the position and orientation of the SWNT with electron beam lithography (Vistec EBPG 5000plus ES). The exposed PMMA was removed with a mixture of methyl isobutyl ketone (MIBK). Finally nickel electrodes (thickness: ~200 nm) were deposited on the samples via thermal evaporator and the SWNT devices can be obtained after lift-off in acetone. The fabrication details of the SWNT devices are shown in Supporting Information S1. The electrical characteristics of the SWNT devices were measured in ambient conditions using Agilent B1500A.

**Switch Frequency.** The switching speed of the SWNT memory device can be estimated using the cylindrical cantilever mode:[31, 32]

$$f = \frac{1}{\sqrt{2}\pi} \frac{r}{L^2} \sqrt{E/\rho}$$

where $E$ is Young's modulus (~1 TPa), $\rho$ is density of SWNT (~1.6 g×cm$^{-3}$), $r$ is diameter of SWNT (1.6 nm), and $L$ is the deflection length (~100 nm). Using this model, the switching frequency is calculated to be 880 MHz.

**Power Consumption.** The SWNT memory device was operated via electrostatic actuation and during the operations the electrostatic energy is transformed into the elastic energy of the deflected SWNT. Thus, very small power was required during the operation. The electrostatic energy can be expressed by the following equation for the capacitive energy: $E_C = 1/2\ CV^2$, where $C$ and $V$ are the capacitance and the voltage between the SWNT cantilever and the source electrode, respectively. Typical value of the capacitor $C$ is estimated to be $3.6 \times 10^{-19}$ F (calculation details in Supporting Information S6). Thus, for a pull-in voltage of 1.5 V, the resultant writing energy of the SWNT memory device is calculated to be $4.1 \times 10^{-19}$ J per bit.

This writing energy is extremely low and several orders of magnitude lower than those reported previously.[19]

ASSOCIATED CONTENT

**Supporting Information**

The Supporting Information is available free of charge on the ACS Publications website.

> The fabrication details of the SWNT devices; The characterization of the suspended section of SWNT; The fabrication and operation of a semiconducting SWNT memory in ambient conditions; Details of the Simmons fitting procedure; Stable SET operations of SWNT memory devices; Calculation of the capacitor made of the SWNT and electrodes.

AUTHOR INFORMATION

**Corresponding Author**

*E-mail: slf@nanoctr.cn

*E-mail: hzhou7@central.uh.edu

*E-mail: wgchu@nanoctr.cn

**Notes**

The authors declare no competing financial interest.

ACKNOWLEDGMENTS

This work was supported by National Science Foundation of China (Grant No. 51472057) and the Major Nanoprojects of Ministry of Science and Technology of China (2016YFA0200403). W.G.Chu acknowledges financial support from the Strategic Priority Research Program of the Chinese Academy of Sciences (Grant XDA09040101).

**Figures and figure captions**

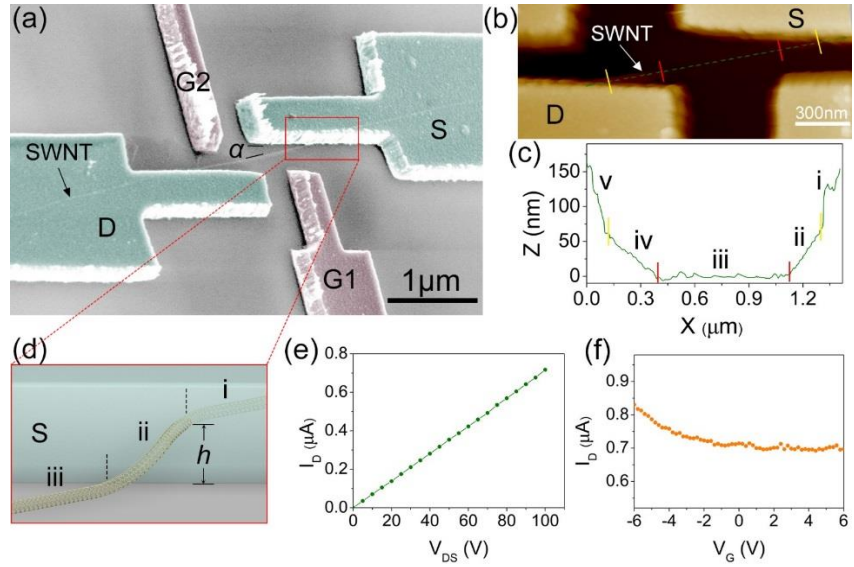

Figure 1. Typical SWNT devices. (a) A false-color SEM image of a SWNT device with a tilted angle of 60°. An individual SWNT is connected to two nickel electrodes: source (S) and drain (D). Two side gates (G1 and G2) are patterned at the nearby location of source and drain electrodes, respectively. There is a small angle (α ~ 10°) between the SWNT and source electrode. (b) AFM topographic image and (c) the relevant height profile of a SWNT device with tapping mode. These results indicate that the SWNT can be divided into five sections: embedded section in the source electrode (section i); suspended and oblique section (Section ii); adhered to the substrate (section iii); suspended and oblique section (section iv); embedded section in the drain electrode (section v). (d) Schematic diagram showing section i, ii, and iii of the SWNT. (e) Typical drain current ($I_D$) versus drain voltage ($V_{DS}$) of a SWNTs device. At low bias, the low resistance (R≈140 KΩ) and linear relationship indicate Ohmic contact between the electrode and the SWNT. (f) Drain current ($I_D$) versus gate voltage $V_G$ (at $V_{DS}$=100 mV) of a typical metallic SWNT device at room temperature. The $V_G$ was applied to the back gate of doped silicon, and the thickness of the $SiO_2$ layer is 300 nm.

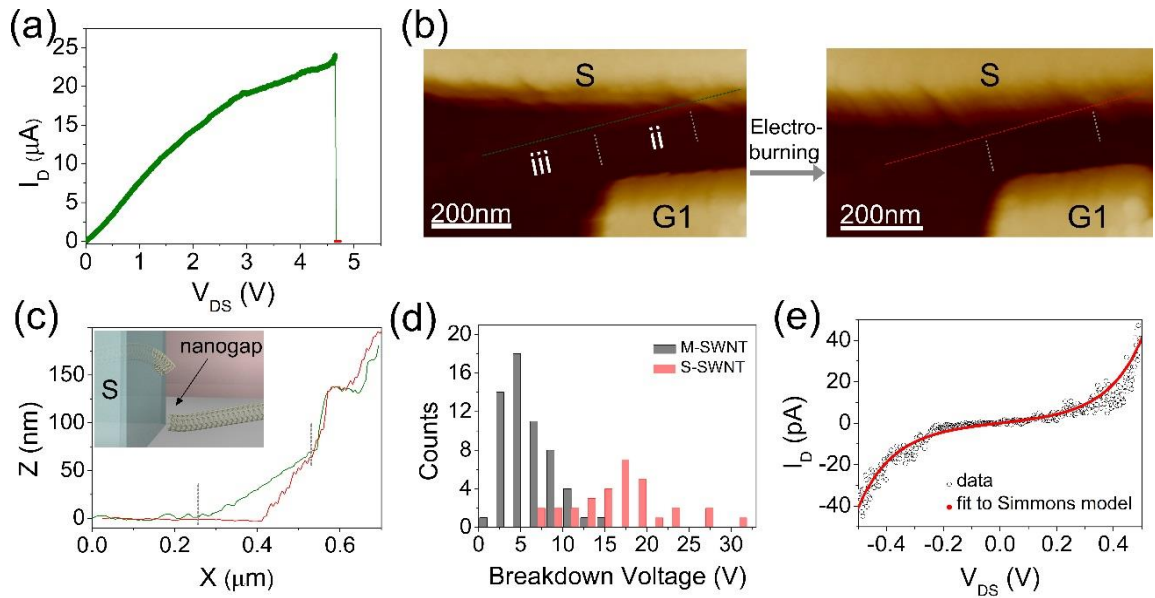

Figure 2. Fabrication of the nanogap using electroburning method. (a) Typical *I-V* curve during electroburning of a SWNT device in ambient conditions. A sharp decrease of current is observed at 4.6 V, indicating electrical breakdown of SWNT (resistance: ~ GΩ). (b) AFM characterizations of the SWNT device before (left) and after (right) electroburning process. (c) The height profile along the SWNT of the AFM images before (green line) and after (red line) electroburning process, respectively. After electroburning, the height of the suspended section (ii) of SWNT becomes smaller. *Inset:* After electroburning, a nanogap was created between the end of the SWNT and the source electrode. (d) Histograms of the electrical breakdown voltage for 58 metallic SWNTs (gray) and 31 semiconducting SWNTs (red) devices. (e) Typical *I-V* trace of a SWNT nanogap. The red line is the fitted curve with the Simmons model, by which the size of the nanogap can be estimated.

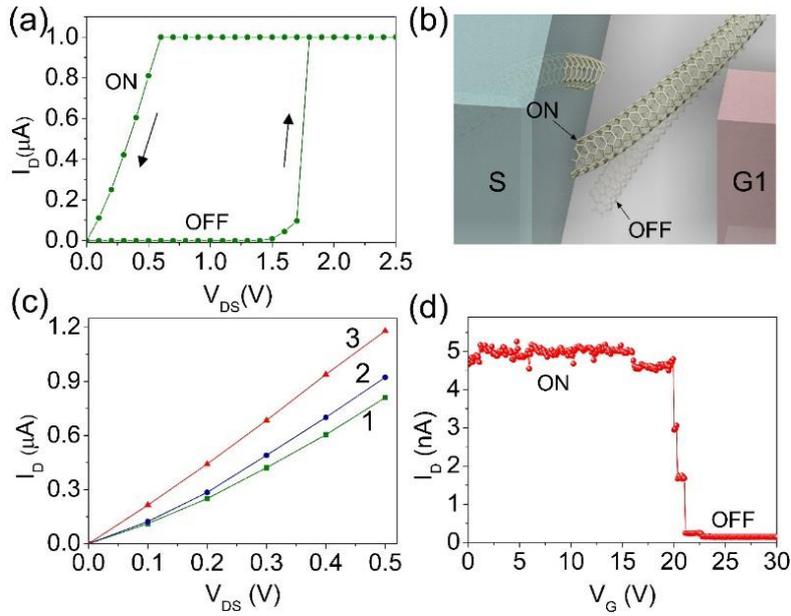

Figure 3. Basic operations of the switch device in ambient conditions. (a) *I-V* characteristics of a SWNT device switching from OFF to ON state. At first, the resistance is high (~GΩ), indicating the state of OFF. When an increasing voltage applies on the source electrode and reaches a critical value (~1.7 V), the resistance of the device becomes low (<1 MΩ), suggesting the state of ON. For all switching experiments, a current compliance of 1 μA was used. (b) Schematic illustrations of the SWNT switch device in the ON (solid) and OFF (translucent) states. (c) Nonvolatility of the switch device. After the SWNT device is switched into ON (marked with "1"), the state can be kept as indicated by the successive sweeping of voltage (marked with "2" and "3"), indicating the nonvolatility of the SWNT device. (d) *I-V* characteristics of the memory device switching from ON to OFF state. An increasing voltage ($V_G$) is applied at the gate electrode (G1) while a small bias (5 mV) is applied between the drain and the source electrodes. The corresponding current $I_D$ is monitored as the $V_G$ is swept up. At first, the resistance of the device is low (<1 MΩ). When the V$_G$ reaches a critical value (switching OFF voltage at 21 V), the resistance becomes very high (~GΩ). As a result, the SWNT device is switched into OFF state.

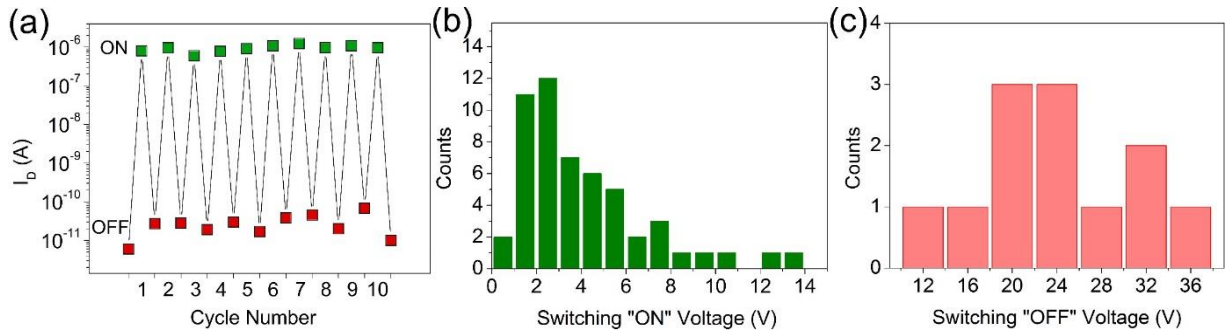

Figure 4. Endurance test of switching operations. (a) Switching between ON and OFF state multiple times for a typical device in ambient condition. The values of drain current ($I_D$) are recorded at the drain voltage ($V_{DS}$) of 500 mV. The ON/OFF ratio of the memory device is about $10^5$. (b) Histogram of switching ON voltage for 53 devices. (c) Histogram of switching OFF voltage for 12 devices.

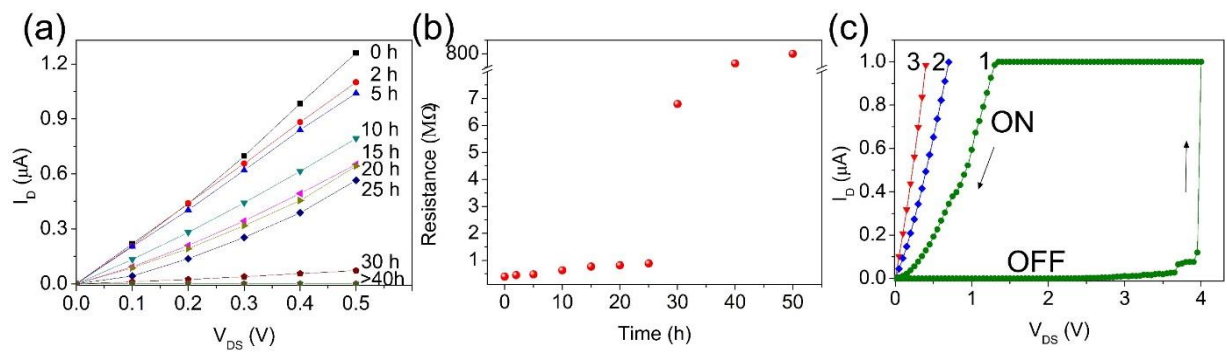

Figure 5. The retention time test in ambient conditions. (a) *I-V* curves versus time at ON state. As the time increases, the device remains at the ON state though the resistance increases gradually. After 40 hours, the resistance of the device becomes high (> 500 MΩ), suggesting the state of OFF. (b) The corresponding evolution of resistance depending on the time at ON state. (c) The re-switching of OFF into ON state. After 40 hours, when the drain voltage is swept up, the state of the device turns into ON at a voltage about 4.0 V. The SWNT device shows nonvolatility as indicated by the subsequent *I-V* curves marked with "2" and "3".